# Large topological Hall effect in a geometrically frustrated kagome magnet $Fe_3Sn_2$

Hang Li,[1,2] Bei Ding,[1,2] Jie Chen,[1,2] Zefang Li,[1,2] Zhipeng Hou,[3] Enke Liu,[1,4] Hongwei Zhang,[1] Xuekui Xi[1], Guangheng Wu,[1] Wenhong Wang[1,4*]

[1]Beijing National Laboratory for Condensed Matter Physics, Institute of Physics, Chinese Academy of Sciences, Beijing 100190, China

[2]University of Chinese Academy of Sciences, Beijing 100049, China

[3]South China Academy of Advanced Optoelectronics, South China Normal University, Guangzhou510006, China

[4] Songshan Lake Materials Laboratory, Dongguan, Guangdong 523808, China

*Corresponding author. Email: wenhong.wang@iphy.ac.cn

**Abstract**

We report on the observation of a large topological Hall effect (THE) over a wide temperature region in a geometrically frustrated $Fe_3Sn_2$ magnet with the kagome-bilayer structure. We found that, the magnitude of the THE resistivity increases with temperature and reaches -0.875μΩ·cm at 380K. Moreover, the critical magnetic fields with the change of THE are consistent with the magnetic structure transformation, which indicates the real-space fictitious magnetic field proportional to the formation of magnetic skyrmions in $Fe_3Sn_2$. The results strongly suggest that the large THE is originated from the topological magnetic spin textures and may open up further research opportunities in exploring emergent phenomena in kagome materials.

The interplay between electron spin and topological magnetic textures, such as magnetic skyrmions[1], has attracted growing interest for its basic scientific importance and the technological applications [2,3]. Recently, magnetic skyrmionic topological spin texture has been discovered in chiral $B$20-type magnets (MnSi[4], $Fe_{1-x}Co_xSi$[5], FeGe[6] and $Cu_2OSeO_3$[7]), centrosymmetric magnets ($La_{0.5}Ba_{0.5}MnO_3$[8], $La_{1.37}Sr_{1.63}Mn_2O_7$[9], MnNiGa[10] and $Fe_3Sn_2$[11]), artificial interface system (Fe/Ir(111)[12], FePd/Ir(111)[13] and Co/Ni/Cu[14]) and Heusler alloys ($Mn_{1.4}PtSn$[15]). According to the previous studies [16-18], the skyrmion spin textures will generate a local spin chirality $\chi = S_i \cdot S_j \times S_k$, where $S_i$, $S_j$ and $S_k$ represent three nearest non-collinear spin as shown in Fig 1.(a), which further induces Berry phase to the wave function of conduction electrons and then contributes to the Hall resistivity $\rho_{xy}$[19]. Since this additional contribution to the Hall effect originates from the topological spin texture, the term topological Hall effect (THE) is coined, which has been widely found in the metallic skyrmion-hosting materials[20-22].

The kagome structure is a hexagonal mesh lattice which is named from the traditional Japanese woven bamboo pattern[23]. Materials with kagome structure have been pursued by researchers for a long time, because they possess many interesting physical phenomena, such as quantum spin liquid[24], topological insulator[25], Dirac[26] or Weyl fermions[27], magnetic skyrmions[11] and so on. In particular, for the geometrically frustrated kagome $Fe_3Sn_2$ magnet, a large anomalous Hall effect[28] and massive Dirac fermions[29] have been reported. In addition, our recent studies on the $Fe_3Sn_2$ have confirmed that the topological spin textures exist over a wide temperature and magnetic field (T-B) region,[11,30] and a many-body spin-orbit tunability emergent at low temperature[31]. However, to confirm the magnetic structure and the origin of the skyrmion spin textures in kagome $Fe_3Sn_2$ magnet, a one-to-one correspondence between magnetic structure transformation and the THE is highly desired. In this letter, we demonstrate skyrmion-derived THE through the combination of magnetic and transport measurements in a geometrically frustrated $Fe_3Sn_2$ magnet with the kagome-bilayer structure. As a result, the maximum magnitude of THE increases with temperature and reaches a large value (-0.875μΩ·cm) at 380K. The variation of magnetic field dependent THE is consistent with the magnetic structure transformation, which indicates the real-space fictitious magnetic field proportional to the formation of topological magnetic spin textures.

Figure 1(a) shows the structure of $Fe_3Sn_2$ lattice (space group $R\bar{3}m$) is composed of Fe-Sn kagome bilayers and Sn layers, which stack along the c-axis alternatively. In the Fe-Sn layers, the Fe atoms form hexagonal meshes and the centers of meshes are occupied by Sn atoms. Note that, there are two different kinds of Fe–Fe bond lengths (2.732Å and 2.582Å) in the Fe-Sn layers and the adjacent Fe-Sn layers present a certain degree of offset. Another Sn sites form six-member rings inserting into the Fe-Sn bilayers. The magnetic moments mainly origin from Fe atoms (~2μ$_B$/Fe) and the nearest three Fe atoms form a frustrate structure[32-34].

Single crystals of $Fe_3Sn_2$ were synthesized by the Sn-flux method with a molar ratio of Fe:Sn = 1:19. Fe (purity 99.95%) and Sn (purity 99.99%) grains were mixed and placed in an alumina crucible, which was sealed in a tantalum tube under partial Argon atmosphere. The tantalum tube was sealed in a quartz tube to avoid oxidation. Then, the quartz tube was placed in a furnace and kept at 1,150 °C about 48 h, and then it was cooled from 910 °C to 800 °C at a rate of 1.5°C/h. Finally, the quartz tube was moved quickly into the centrifuge to separate the excess Sn flux. As shown in the Fig. 1(b), the shape of the $Fe_3Sn_2$ single crystal presents hexagonal with shiny surface. X-ray diffraction and the selected-area diffraction pattern was taken at room temperature by Bruker D2 x-ray machine with Cu K$_α$ radiation (λ = 1.5418Å) and scanning transmission microscopy (Titan G2 60-300, FEI), respectively. These results strongly indicate that the hexagonal plane is ab-plane.

The magnetization and transport properties were performed in the physical property measuring system (PPMS, Quantum Design). The demagnetization factor $N_d$ = 0.65 has been taken into consideration in data processing ($B = \mu_0[H + (1 − N_d)M]$). The magnetization curves were measured from 5K to 380K at the external field which is vertical and parallel to ab-plane (See Fig. S1 of supplementary material), respectively.

When the external field **B** is vertical to the ab-plane, as shown in Fig. 1(c), there is a clear cross between high and low temperatures magnetization curves at low field region (other detailed data were shown in Fig. S2 of supplementary material). The slopes of magnetization curves at low field with various temperature are shown in inset of Fig. 1(c), the change of the slopes becomes much steeper at temperatures below 100K, which reaches the maximum at 100K. The temperature dependent magnetic anisotropy constant $K_u$ is presented in Fig. 1(d), a turning point appears clearly at 150K. These abnormal changes may arise from the rotation of easy magnetization axis from c-axis to ab-plane with the decrease of temperature [34]. However, we should point out that, as shown in inset of Fig. 1(d), the moment still favor ab-plane at 380K, which is different from what has been reported in previous literature[32].

The longitudinal resistivity $\rho_{xx}$ and Hall resistivity $\rho_{xy}$ were measured simultaneously by standard six-probe method from 5K to 380K as shown in Fig. 2. The sample was assembled on a puck with epoxy and connected to the measuring circuit by silver epoxy and platinum wires (0.02mm in diameter). From the measurement of resistivity as a function of temperature (See Fig. S3 of supplementary material, sample A), a very large residual resistivity ratio (RRR = $\rho_{xx}(300K)/\rho_{xx}(5K)$), as high as 58.8, was found in our $Fe_3Sn_2$ single crystals, which indicates the high quality of the sample. To get rid of the influence of voltage probe misalignment, we processed the data by the formulas of $\rho_{xx} = [\rho_{xx}(B) + \rho_{xx}(-B)]/2$ and $\rho_{xy} = [\rho_{xy}(B) - \rho_{xy}(-B)]/2$, respectively. The data of Magnetoresistance (MR = $(\rho_{xx}(B) - \rho_{xx}(0))/\rho_{xx}(0)$) are shown in Fig. 2 (a), we can see clearly that the sign changes of MR at T = 120K. When T < 120K the sign of MR is positive and the value increases rapidly with the decrease of temperature. However, when T > 120K, the sign is negative and the absolute value increases slightly with temperature. As shown in Fig. 2 (b), the value of $\rho_{xy}$ as a function of temperature, it is shown clearly that almost no anomalous Hall effect was observed below 80K. The sign change of MR and $\rho_{xy}$ may also result from the rotation of easy axis with temperature[34], which also are consistent with previous reports[28]. However, the discrepancy in transition temperature between MR and $\rho_{xy}$ may arise from their difference sensitivity to the rotation of easy axis, which needs further study.

The inset of Fig. 3(a) presents the fitted results of $\rho_{xy}$ with standard formula $\rho_{xy} = R_0 B + S_A \rho_{xx}^2 M$, where $R_0$ is the ordinary Hall coefficient, $S_A$ is the scaling coefficient independent of temperature, and M is the magnetization[35]. It is shown clearly that the fitting curve does not coincide completely with observed data at low field region, indicating there are other contributions to $\rho_{xy}$. On the other hand, we subtracted fitting data from observed data and then obtained a peak with negative values at low field region. The extracted values ($\rho_{xy}^T$) at various temperature are shown in Fig. 3 (a), we can see that the peaks only exist above 120K and the absolute values of the peaks increase with temperature. The maximum of the absolute value reaches 0.875μΩ·cm at 380K.

It is well known that Fe$_3$Sn$_2$ is a geometrically frustrated magnet[34] and has diverse topological magnetic domain above 130K[11,30]. To understand the relation between the peaks and the magnetic structure of Fe$_3$Sn$_2$, we extracted the critical magnetic field corresponding to the maximum absolute value ($B_{peak}$) and vanishing ($B_0$) of the peaks at various temperature and compared with the phase diagram of magnetic structure from the previous report[11]. It is shown clearly that the $B_{peak}$ closes to the conversion region of trivial bubbles to skyrmionic bubbles, and the $B_0$ almost consists with the critical field of skyrmionic bubbles disappearance. Firstly, the values of $B_{peak}$ keep almost unchanged by varying temperature, which also implies topological protection in Fe$_3$Sn$_2$. Secondly, the size of topological magnetic domain[11] (~150nm) is much larger than the crystallographic lattice (a=b=0.53074nm, c=1.97011nm), we can therefore exclude the influence from non-trivial geometry in crystallographic lattice[17]. Thirdly, the magnetic structure completely transform into spin glass state within ab-plane at T < 120K, the contribution to Hall resistivity from local spin chirality $\chi$ is negligible due to the disorder of spin[34]. Thus, there is almost no $\rho_{xy}^T$ at T < 120K. Based on the above analysis, we can confirm that the peak originates from the topological magnetic structure of $Fe_3Sn_2$.

The negative values of $\rho_{xy}^T$ is due to the carriers are electrons in $Fe_3Sn_2$, thus the sign of $R_0$ is negative which is consistent with the $\rho_{xy}^T$[20]. Moreover, there are non-zero values of $\rho_{xy}^T$ even the external field almost approach zero, which is owing to the intrinsic frustrated magnetic structure. Compared with other skyrmion-hosting materials as shown in Fig. 4, we can predict a much larger THE may be in $Fe_3Sn_2$ if one can measure at higher temperatures, because the skyrmions are still stable even at 630K[30].

In conclusion, we report a large THE which is induced by the local spin charily from topological magnetic structure in $Fe_3Sn_2$ over a wide temperature region. The sign of THE is negative and the absolute value increases with the temperature. The critical magnetic field of THE is consistent with the magnetic structure transformation, which reflects the real-space fictitious magnetic field proportional to the formation of topological magnetic spin textures. The sign change of MR and Hall effect may attribute to the easy axis variation with temperature at temperatures below 120K, although the detailed evolution of microscopic magnetic structures as function of temperature needs further investigations.

**Supplementary Material**

● S1: Magnetization curves measured at **B** parallel to ab-plane and c-axis, respectively.

● S2: Magnetization curves measured at **B** parallel to c-axis.

● S3: Resistivity of different samples varied with temperature at current along ab-plane and zero magnetic field.

● S4: Magnetoresistance and Hall resistivity from the sample B and C, which consistent with the transition temperature of sample A.

● S5: The I-V curves of the sample measured at different temperature and zero magnetic field.

**ACKNOWLEDGMENT**

This work was supported by the National Key R&D Program of China (Grant Nos. 2017YFA0206303, 2017YFA0303202), National Natural Science Foundation of China (Grant No. 11574137), and the Key Research Program of the Chinese Academy of Sciences, KJZD-SW-M01.

**Figure captions**

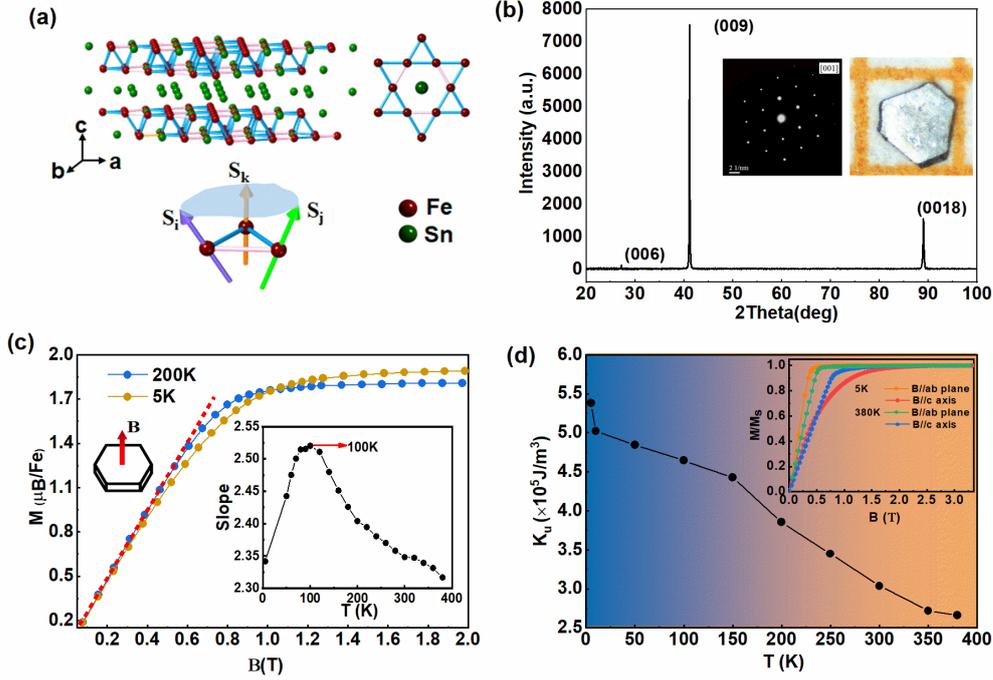

**FIG. 1** (color online). **(a)** Schematic of bilayer kagome lattice $Fe_3Sn_2$, which the Fe-Sn layers and Sn layers stack along c axis alternatively. **(b)** The X-ray diffraction pattern of the hexagonal plane. Inset shows the photo of single crystal and selected-area diffraction pattern taken by scanning transmission microscopy. **(c)** The magnetization curves at 5K and 200K with magnetic field **B** along c-axis. Inset shows the slopes of the low field magnetization curves at various temperature. **(d)** The anisotropy constant $K_u$ varies with temperature. Inset shows the magnetization curves at 5K and 380K with B parallel to ab-plane and c-axis, respectively.

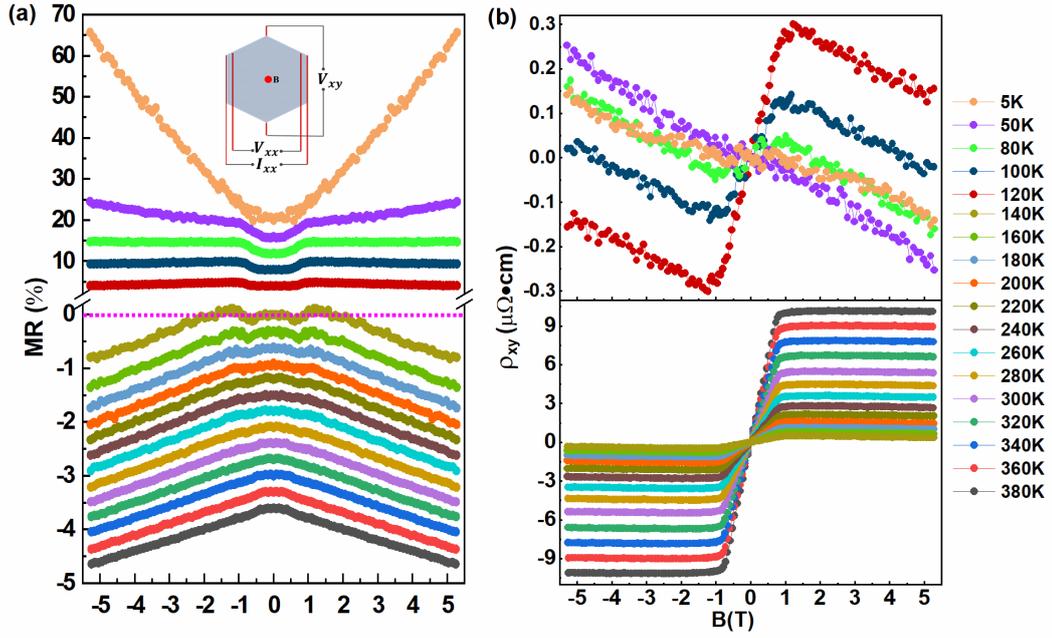

**FIG. 2** (color online). **(a)** Magnetoresistance and **(b)** Hall resistivity from the same sample measured simultaneously at various temperatures. The inset **(a)** shows the schematic of wire arrangement for the measurements. The pink dotted line represent the zero of MR. The curves shift a little to make it clear. The upper plane of **(b)** shows the Hall resistivity below 120K and the lower plane **(b)** shows the Hall resistivity between 140 and 380K.

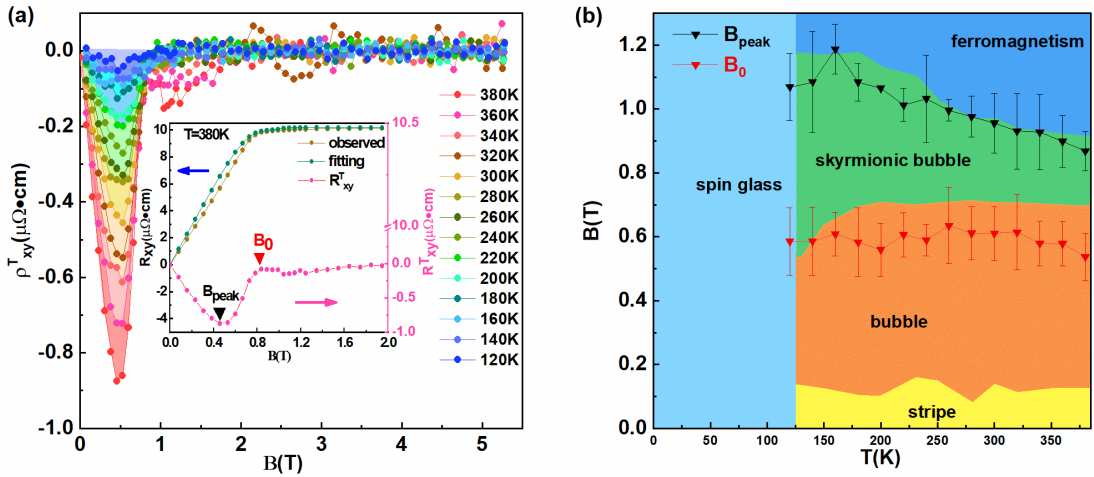

**FIG. 3** (color online). Topological Hall resistivity at different temperature and magnetic field. **(a)** Extracted topological Hall resistivity above 120K. The inset shows the process of extracting. The black and red triangles correspond to the maximum absolute value ($B_{peak}$) and vanishing ($B_0$) of the peaks. **(b)** Comparison of the magnetic phase diagram and the critical magnetic fields of THE. The error bars were added based on the results measured on three different samples (A, B, and C).

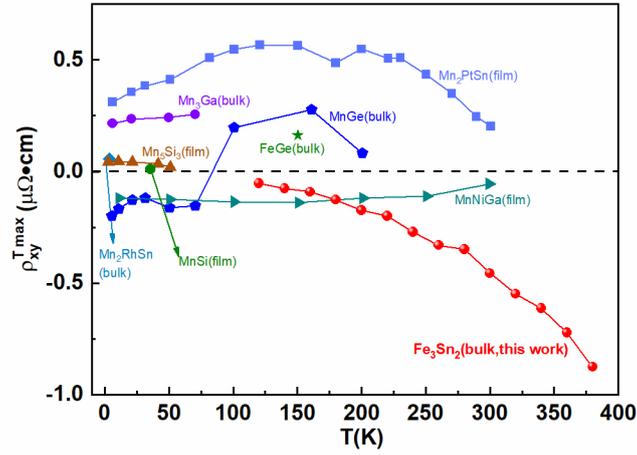

**FIG. 4** (color online). Comparison of topological Hall values of various skyrmion materials. The black dotted line represent the zero position. The data from Refs.[21,35-41]

# Supplementary material

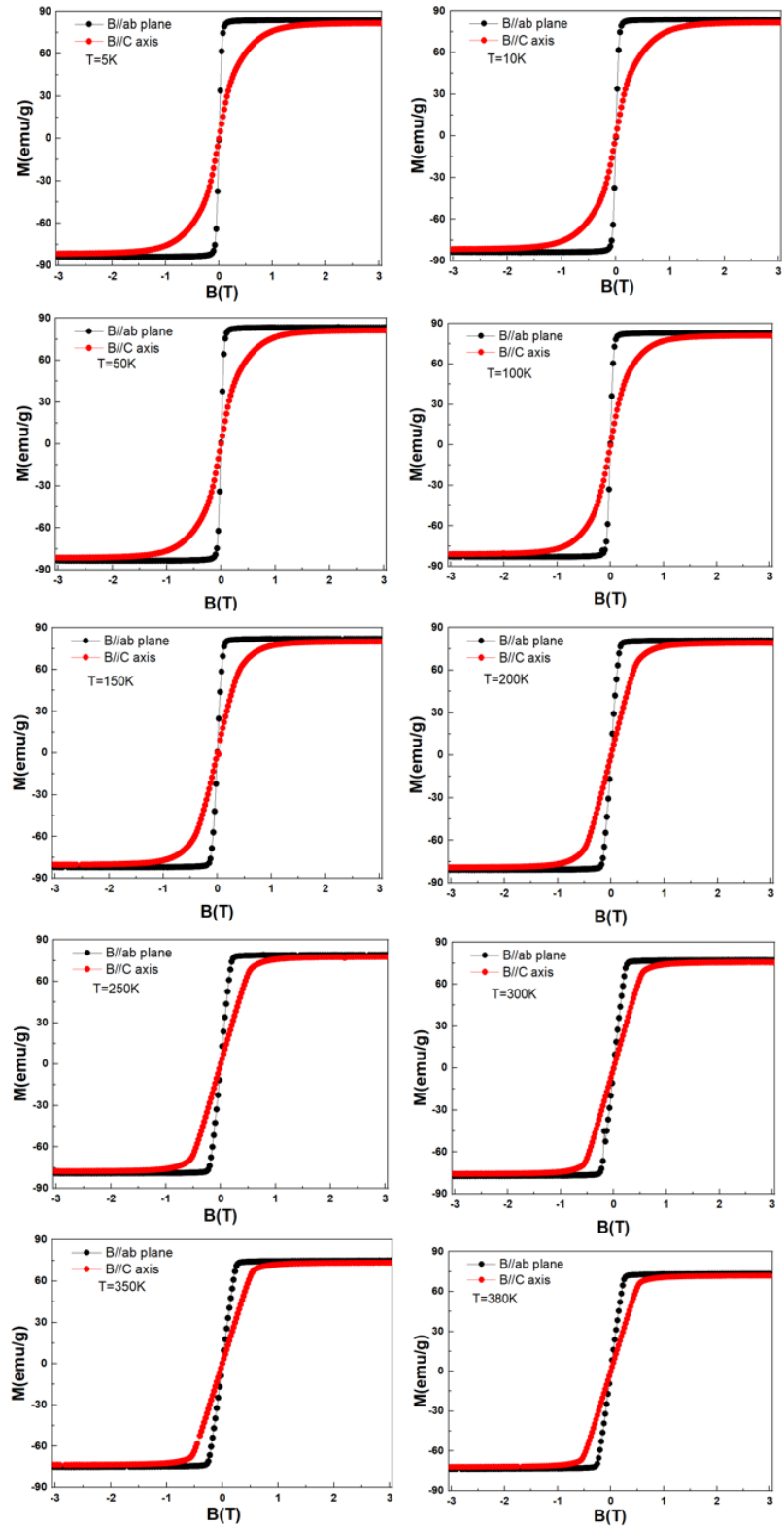

**Fig.S1** Magnetization curves measured at **B** parallel to ab-plane and c-axis, respectively. The data indicates that the ab-plane always more easily magnetize than c-axis from 5K to 380K. However, it is clearly shown that the difference between them decreases as the rise of temperature, which suggests that the easy magnetization direction gradually rotates to the c axis.

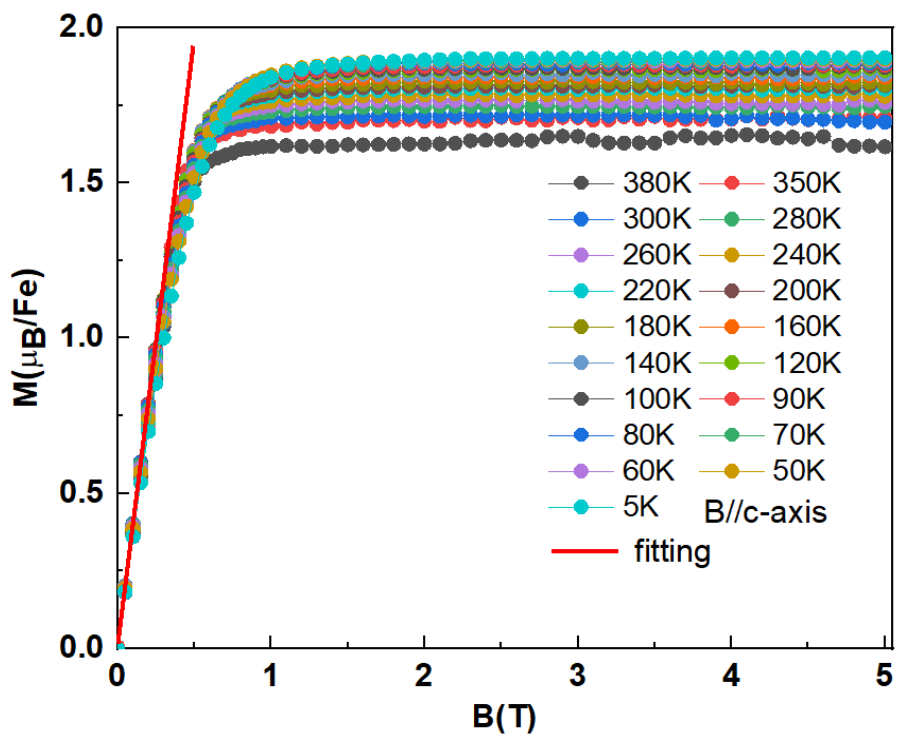

**Fig.S2** Magnetization curves measured at B parallel to c-axis. It is clearly shown there is a cross at low field region among different temperature. The red line represents the fitting magnetization line at low field.

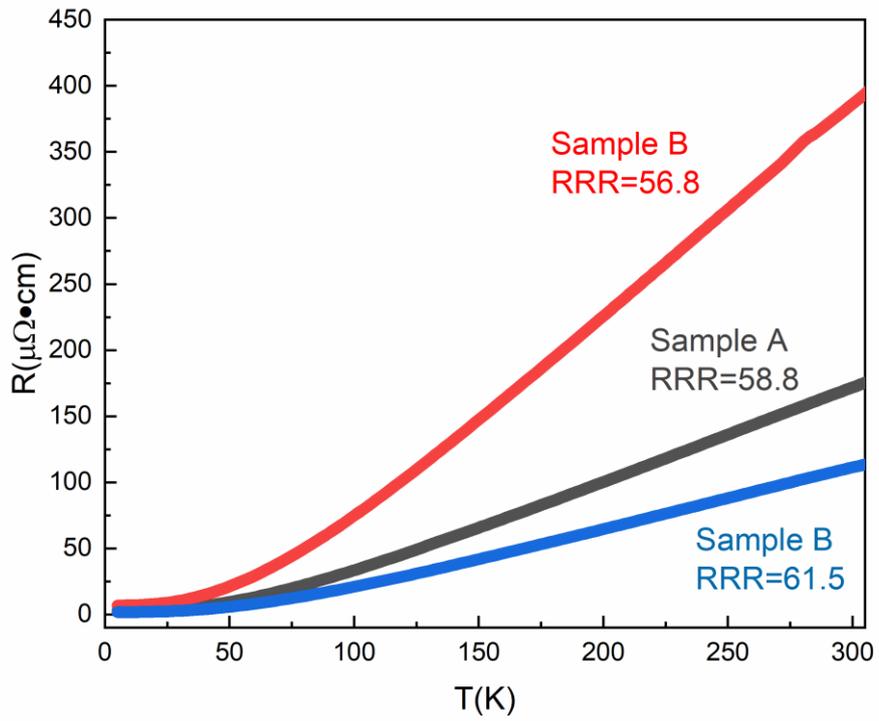

**Fig. S3** Resistivity of different samples varies with temperature at current along ab-plane and zero magnetic field. This is a typical metallic curve and the high RRR indicates the high quality of the signal crystal samples.

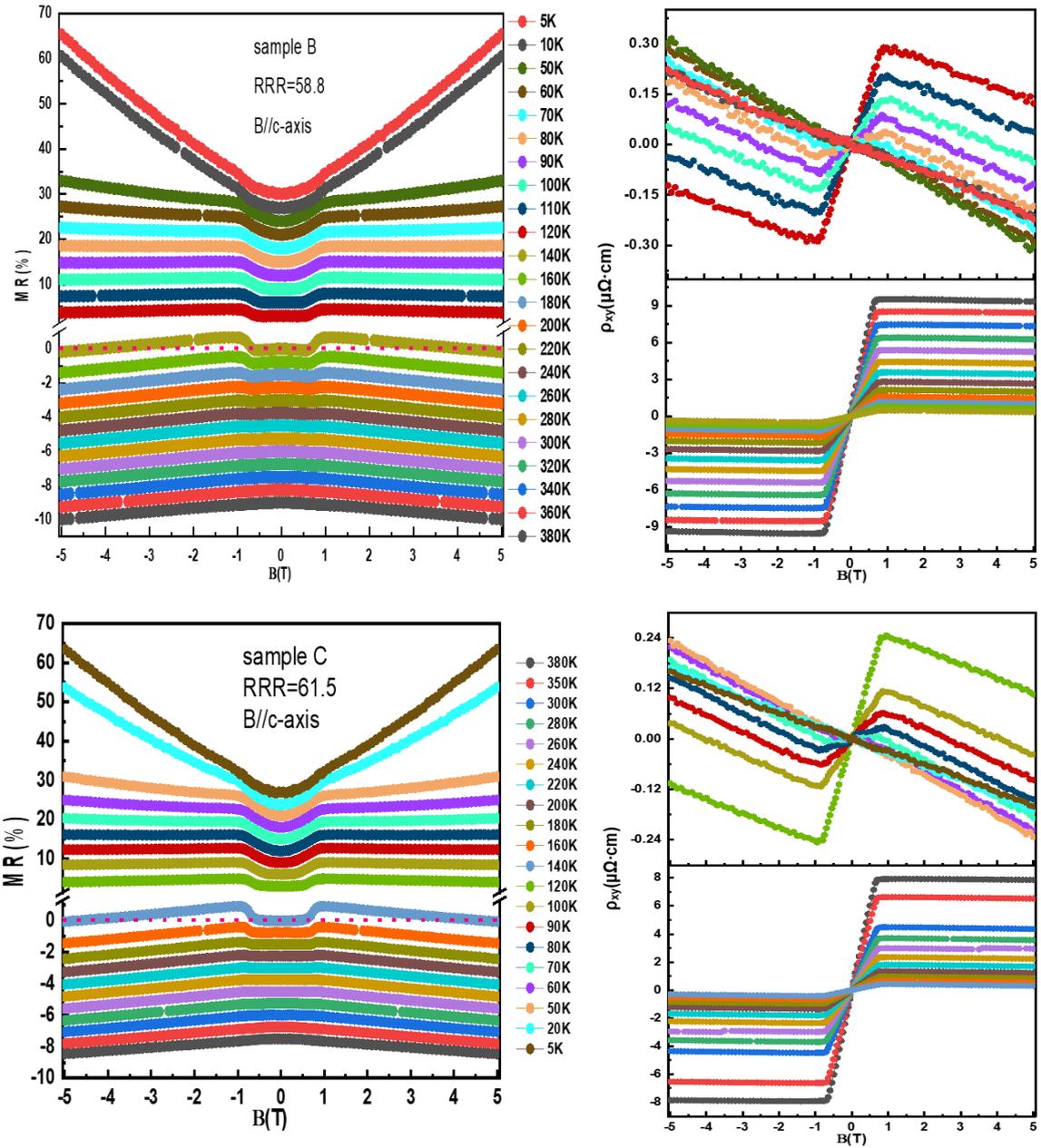

**Fig. S4** Magnetoresistance and Hall resistivity from the sample B and C, which are consistent with the transition temperature of sample A.

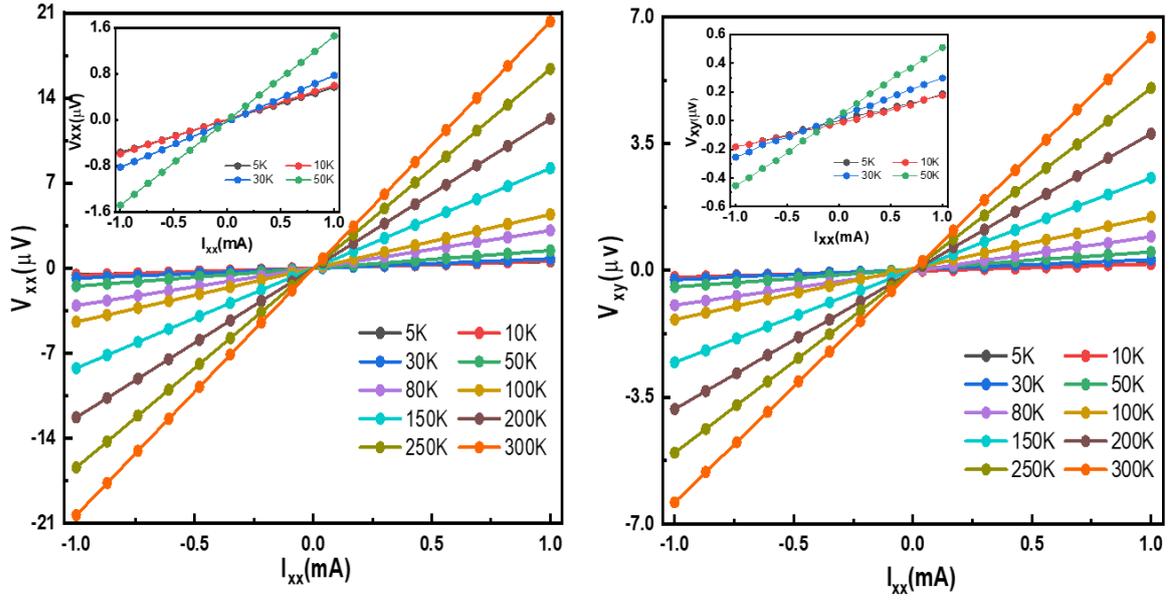

**Fig. S5** The I-V curves of the sample A measured at different temperature and zero magnetic field. The good linear fitting of the data implies the good ohmic contact of the sample.